\documentclass{article}
\pdfoutput=1

\usepackage[margin=1in]{geometry}
\usepackage{graphicx}
\usepackage{epstopdf}
\usepackage{cite}
\usepackage{comment}
\usepackage{amsmath,tabu,amssymb,amsfonts,varwidth}
\usepackage{xcolor}

\usepackage[affil-it]{authblk}

\newcommand{\me}{\mathrm{e}}

\newcommand{\VE}{\mathbf}
\newcommand{\DE}{\mathrm{d}}

\newcommand{\pder}[2]{\frac{\partial#1}{\partial#2}}

\newcommand{\Avg}[1]{\left<#1\right>}

\newcommand{\iinteg}[3]{\iint_{#1}\!{#2}\,\DE #3}
\newcommand{\dinteg}[4]{\int_{#1}^{#2}\!{#3}\,\DE #4}

\begin{document}


\title{Three-dimensional contact of TITH cartilage layers, a closed-form solution}

\author{Gennaro Vitucci%
  \thanks{Corresponding author: \texttt{gev4@aber.ac.uk}}}
\author{Gennady Mishuris}
\affil{Department of Mathematics, IMPACS, Aberystwyth University, Ceredigion, SY23~3BZ, UK}

\date{}
\maketitle
\section*{Abstract}
Inhomogeneity and anisotropy play a crucial role in attributing articular cartilage its properties. The frictionless contact model constructed here consists in two thin biphasic transversely isotropic transversely homogeneous (TITH) cartilage layers firmly attached onto rigid substrates and shaped as elliptic paraboloids of different radii. Using asymptotic techniques, a solution to the deformation problem of such material has been recently obtained extending previous ones referred to homogeneous materials. The layer itself is thin in comparison with the size of the contact area and the observed time is shorter than the hydrogel characteristic time. The emerging three-dimensional contact problem is solved in closed-form and numerical benchmarks for constant and oscillating loads are given. The results are shown in terms of contact pressure and approach of the bones. The latter is derived to be directly proportional to the contact area. Existing experimental data are reinterpreted in view of the current model formulation. Comparisons are made with existing solutions for homogeneous biphasic materials in order to underline the functional importance of inhomogeneity in spreading the contact pressure distribution across the contact area. Particular attention is paid to the applicability of the retrieved formulas for interpreting measurements of \textit{in vivo} experiments. Future directions are also prospected. 


\section{Introduction}\label{sec:intro}
Articular cartilage covers the bones extremities converging into the diathrodial joints. It performs the task of improving the load transmission cutting down friction and stress peaks. This biological tissue peculiar properties are enhanced by a complex multiphasic structure. The solid phase mainly consists of a porous proteoglycan matrix reinforced by collagen fibers. Their inhomogeneous arrangement across the layer depth causes inhomogeneity and anisotropy both in the stiffness and permeability of the solid skeleton. The voids are saturated an by interstitial fluid which is chiefly composed of water and mobile ions causing electro-chemo-mechanical interactions (e.g. \cite{lai1991triphasic,loret2007articular}). Understanding the behavior of such an intricate system has long stimulated scientific research because of the necessity of patient specific diagnosis of degenerative pathologies, such as osteoarthritis, and challenging tissue engineering for adequate replacement (e.g. see \cite{ateshian2015toward,hollister2005porous} for literature review).

A steady progress in computational power encouraged to build biphasic and triphasic fiber-reinforced material models and to search for solutions by use of finite element analysis (\cite{li1999nonlinear,korhonen2003fibril,placidi2008variational,gorke2012consistent}). The correspondence between triphasic and biphasic models and the possible occurring difficulties have been discussed in \cite{ateshian2004correspondence,meng2017effect}. The thinness of the cartilage layers with respect to the size of the bones and contact area, though, may give origin to ill-conditioning, numerical instability and high computational costs due to the necessity for highly refined meshes in the vicinity of the layer (\cite{wilson2005role}). Because of this, analytical formulations still benefit of popularity in the field and are, so far, able to include a wide range of nonlinear effects such as strain-dependency of the material properties and tension-compression nonlinearity (e.g. \cite{mow1980biphasic,soltz2000conewise,holzapfel2015tension}). 

The present work inserts in the discussion about how to analytically solve the contact problem of two biphasic layers attached onto rigid substrates. It is done using an asymptotic approach which enables to retrieve closed-form solutions with the advantage of easily analyzable formulas (\cite{argatov2016rev}). The studies published so far attain to the cartilagineous material modeled first as isotropic homogeneous (\cite{ateshian1994asymptotic}), later as homogeneous but transversely isotropic (\cite{argatovcontact}). Speaking of the utilized geometry, the solution provided in \cite{ateshian1994asymptotic} regarded identical spherical surfaces and it was extended to two different radii spheres in \cite{wu1996modeling,wu1997improved}. A new progress was aimed in \cite{argatov2011elliptical} by the introduction of elliptic paraboloids resulting in elliptical contact areas. Nevertheless the importance of inhomogeneity in the material property distribution across the thickness has been widely explored as a crucial factor in improving superficial fluid support, thus protecting the tissue from damage (\cite{krishnan2003inhomogeneous,federico2008anisotropy}). This was the reason for our recent study \cite{vitucci2016}, summarized in Sec.~\ref{sec:model}, where a special exponential-type inhomogeneity was introduced. It provided, to the best of our knowledge, the first such asymptotic solution to the deformation of an inhomogeneous biphasic layer, whereas studies existed already concerning monophasic layers obtained in the framework of functionally graded materials (\cite{chidlow2013two,tokovyy2015analytical} and literature survey there). 

The solution to the contact problem is derived in Sec.\ref{sec:ansol} and some numerical benchmarks are illustrated in Sec.\ref{sec:numben}. The physical bounds for the model parameters are discussed. Geometry, solicitations and material stiffness and permeability are assigned trying to be as realistic as possible in the framework of the model by ample use of available publications. Two load conditions are exemplified, a constant load and a sinusoidal one. In particular, by means of the retrieved formulas, the utilized contact radii are extracted from the experimental measurements on human tibiofemoral joints provided \cite{hosseini2010vivo}. In Sec.\ref{sec:disc} we draw our conclusions on some aspects which suggest how inhomogeneity turns favorable for this specific biological tissue and on the applicability and limitations of the current model. The need for data which can reveal crucial for mechanics scientists in order to provide effective diagnosis tools are also remarked.

\section{Model and statement of the contact problem}
\label{sec:model}
In the recent work \cite{vitucci2016}, the deformation problem for a thin biphasic transversely isotropic, transversely homogeneous (TITH) biphasic layer was studied. An infinitely extended thin layer, firmly attached along one face, was loaded perpendicularly to the opposite one. The fluid flow was constrained by the two layer faces. The solid matrix was considered linear elastic and the interstitial fluid inviscid, given that the low permeability causes the friction drag to be dominant with respect to the viscous flow. The solid matrix constitutive law is described by the stiffness matrix
\begin{equation}
\VE{A}(z)=\left[\begin{array}{c c c c c c}
A_{11}&A_{12}&A_{13}&&&\\ A_{12}&A_{11}&A_{13}&&&\\ A_{13}&A_{13}&A_{33}&&&\\
&&&2A_{44}&&\\ &&&&2A_{44}&\\ &&&&&2A_{66}\\	
\end{array}\right],
\label{eq:amatrix}
\end{equation}
whose components vary through the local depth-coordinate $z\in[0,1]$ from the surface to the substrate. Also the diagonal permeability tensor was considered TITH of components $\mathrm{diag}(\VE{K}(z))=[K_1,K_1,K_3]$. A special exponential inhomogeneity was allowed:
\begin{equation}\begin{aligned}
&A_{33}=a_{33}\me^{2\gamma z},\quad A_{44}=a_{44}\me^{\alpha z},\quad A_{13}=a_{13}\me^{\alpha_{13}z},\\
&K_3=k_3\me^{-2\gamma z},\quad K_1=k_1\me^{-\gamma_1 z}.
\end{aligned} 
\label{eq:paramdef}
\end{equation}
According to it, in spite of an arbitrary exponential variation of every component, $A_{33}$ and $K_3$ are linked through $\gamma>0$, thus let respectively increase and decrease of the same ratio across the thickness. The derived relation between the contact pressure $P$ and the surface lowering of the layer surface is
\begin{gather}
w
=\bar{\alpha}_0\Delta P
+\bar{\alpha}_1\dinteg{0}{t}{\me^{\bar{\beta}_1(t-\theta)}\Delta P}{\theta}
+\bar{\alpha}_2\dinteg{0}{t}{\me^{\bar{\beta}_2(t-\theta)}\Delta P}{\theta}
+\bar{\alpha}_3 \dinteg{0}{t}{\Delta P(\theta)}{\theta},
\label{eq:disp_i}
\end{gather}
where the operator $\Delta$ represents the Laplacian in the plane orthogonal to $z$. The expression of the coefficients in Eq.~\eqref{eq:disp_i} as functions of the TITH biphasic material parameters of Eq.~\eqref{eq:paramdef} are displayed in Tab.~\ref{tab:albes}.
\begin{table}[ht]
\centering
$\extrarowsep=10pt\begin{tabu}{|c|c||c|c|}
\hline  
\bar{\alpha}_0&\dfrac{2-\me^{-\alpha}(\alpha^2+2\alpha+2)}{\alpha^3a_{44}}h^3&\bar{\alpha}_3&\dfrac{1-\me^{-\gamma_1}}{\gamma_1}h k_1 \\ \hline
\bar{\alpha}_1&\dfrac{a_{13}(\alpha_{13}-\alpha)(1-\me^{\alpha_{13}-\alpha-2\gamma})}{\alpha a_{44}}h k_3&\bar{\beta}_1& (\alpha_{13}-\alpha)(\alpha_{13}-\alpha-2\gamma)\dfrac{a_{33}k_3}{h^2}\\ \hline
\bar{\alpha}_2&(\alpha-2\gamma)\dfrac{1-\me^{-\alpha}}{\alpha a_{44}}h k_3 a_{33}&\bar{\beta}_2&\alpha(\alpha-2\gamma)\dfrac{a_{33}k_3}{h^2} \\ \hline
\end{tabu}$
\label{tab:albes}\caption{Coefficients of the pressure-displacement asymptotic relation in Eq.~\eqref{eq:disp_i} as functions of the material parameters of Eq.~\eqref{eq:paramdef} as derived in \cite{vitucci2016}.}
\end{table}
Such closed-form asymptotic solution was obtained under the conditions that the characteristic scale of the phenomenon along the tissue was much bigger than the thickness $h$ itself and that the observed time $t$ was relatively smaller than the hydrogel characteristic time $\tau_{\text{gel}}=h^2/(A_{33}K_3)$. 

\begin{figure}[ht]
\centering
\includegraphics[width=0.5\textwidth]{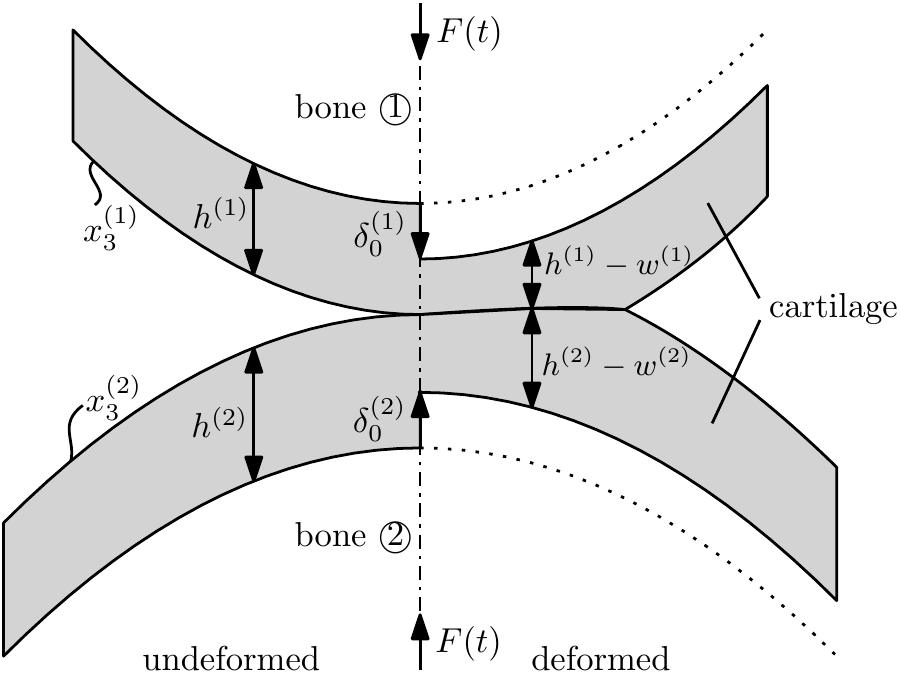}
\caption{Schematic cross section of the contact zone between two bone heads covered by constant thickness cartilage (grey). The two sides of the drawing illustrate the geometry before and after the compression caused by the force $F(t)$.}
\label{fig:contactscheme}
\end{figure}
The articular cartilage joint is the zone where two bone heads $(1)$ and $(2)$, coated by thin films of cartilaginous tissue, get in reciprocal contact. Fig.~\ref{fig:contactscheme} gives an idea of the geometrical changes due to deformation through a finite cross section of the infinitely extended three-dimensional model. The surface displacements $w^{(i)}$ are taken positive if directed toward the respective bones. The bones approach is $\delta_0=\delta_0^{(1)}+\delta_0^{(2)}$. If the two cartilage layers present constant thickness, the problem is stated as
\begin{equation}
\delta_0-w^{(1)}-w^{(2)}=x_3^{(1)}-x_3^{(2)}
,\label{eq:cont0}
\end{equation} 
where the two undeformed surfaces are elliptic paraboloids of equation
\begin{gather}
x_3^{(i)}(x_1,x_2)=\left(\frac{x_1^2}{2R_1^{(i)}}+\frac{x_2^2}{2R_2^{(i)}}\right)(-1)^{(i+1)}.
\label{eq:parabi}
\end{gather}
This way the right-hand side of Eq.\eqref{eq:cont0} may be grouped via a function of the planar coordinates only
\begin{equation}
\Phi(x_1,x_2)=\frac{x_1^2}{2 R_1}+\frac{x_2^2}{2 R_1}
\label{eq:parab}
\end{equation}
expressed through the harmonic averages of the radii 
\begin{equation}\frac{1}{R_k}=\frac{1}{R_k^{(1)}}+\frac{1}{R_k^{(2)}}>0,\end{equation}
being strictly positive. Later on we associate the index $(1)$ to the convex body lying in the upper half-space $x_3>0$.
The studies \cite{argatov2010axisymmetric,koroleva2015analysis} offered a solution to the contact problem which accounts also for the displacement component which is parallel to the contact surface. Despite an increase in computational efforts and loss of simplicity in the obtained formulas, such rigor did not seem to alter dramatically the quantitative results within the usual values of the material in exam and it is consequently neglected in the present work. 
Substituting the pressure-displacement relation Eq.\eqref{eq:disp_i} into Eq.\eqref{eq:cont0} and multiplying both sides by $m=-(\bar{\alpha}_0^{(1)}+\bar{\alpha}_0^{(2)})^{-1}$ leads to
\begin{equation}
\Delta P(t)
+\sum_{j=1}^4\alpha_j\dinteg{0}{t}{\me^{\beta_j(t-\theta)}\Delta P(\theta)}{\theta}
+\alpha_5 \dinteg{0}{t}{\Delta P(\theta)}{\theta}= m(\Phi-\delta_0),\label{eq:cont1}
\end{equation}
once defined $\alpha_5=-m(\alpha_3^{(1)}+\alpha_3^{(2)})$, $\alpha_j=-m\bar{\alpha}_k^{(i)}$ and the exponents $\beta_j=\bar{\beta}_k^{(i)}$ by re-indexing $j=i+2k-2$ for $k=1,2$. It turns useful to introduce the operator $\mathcal{G}$ as 
\begin{equation}
\mathcal{G}y(t)=Y(t)=y(t)
+\sum_{j=1}^4\alpha_j\dinteg{0}{t}{\me^{\beta_j(t-\theta)}y(\theta)}{\theta}
+\alpha_5 \dinteg{0}{t}{y(\theta)}{\theta}.\label{eq:defG}
\end{equation}
In view of Eq.~\eqref{eq:defG}, Eq.~\eqref{eq:cont1} appears now concisely as
\begin{equation}
\mathcal{G}\Delta P(x_1,x_2,t)= m(\Phi(x_1,x_2)-\delta_0(t)). \label{eq:cont2}
\end{equation}
The pressure is set to zero outside the contact area. In the case of cartilage it has been shown that in the superficial area the load is borne mainly by the fluid pressure (\cite{ateshian1994asymptotic,wu2000pressure,argatovcontact}), indeed shear strains are absent because of the absence of friction. It means nullifying also the normal derivative of the pressure at the border $\Gamma(t)$ of $\omega(t)$ and outside:
\begin{equation}
P=0,\qquad\pder{P}{n}=0\qquad \text{on}\quad\Gamma(t)\cup\mathbb{R}^2\setminus \omega(t).
\label{eq:BCs}
\end{equation}
The total external force $F(t)$ is transmitted through the joint which must be balanced on both the cartilaginous surfaces by means of the pressure. Specifically:
\begin{equation}
\iinteg{\omega(t)}{P(x_1,x_2,t)}{\omega}=F(t).\label{eq:equil}
\end{equation}

\section{Analytical solution}
\label{sec:ansol}
The expected contact area between two elliptic paraboloids Eq.\eqref{eq:parabi} of coinciding principal directions, the Cartesian axes $x_1=0$ and $x_2=0$, is elliptical with the border description
\begin{equation}
\Gamma(t): \quad \frac{x_1^2}{a^2(t)}+\frac{x_2^2}{b^2(t)}=1.
\end{equation}
Consequently, adopting a similar line of reasoning as \cite{argatov2010closed}, the solution to Eq.~\eqref{eq:cont2} is searched in the following form: we assume that $\mathcal{G} P(t)$ may be expressed through the auxiliary variable $p(t)$; then we factorize $p(x_1,x_2,t)$ is the form of a product of a time function $p_0(t)$ and a part which fulfills \textit{a priori} the boundary conditions Eqs.~\eqref{eq:BCs} on $\Gamma(t)$. Naming $a(t)$ and $b(t)$ respectively the major and minor semi-axes of the elliptical contact area to determine,
\begin{equation}
p=\mathcal{G}P(x_1,x_2,t)=p_0(t)\left(1-\frac{x_1^2}{a^2(t)}-\frac{x_2^2}{b^2(t)}\right)^2,
\label{eq:ppoft}
\end{equation}
which transforms to the problem Eq.~\eqref{eq:cont2} into
\begin{equation}
\Delta p=m(\Phi-\delta_0). 
\label{eq:probdp}
\end{equation}
Substituting Eq.~\eqref{eq:ppoft} into Eq.~\eqref{eq:probdp}, the resulting relation can be split into three simultaneous conditions by equating the coefficients of $x_1^2$, $x_1^2$ and constant terms with respect to the planar coordinates. It will be soon evident how convenient it is to introduce the ellipse aspect ratio $s(t)=b(t)/a(t)$.
\begin{equation}\left\{
\begin{aligned}
&\frac{4 p_0}{a^4}\frac{3s^2+1}{s^2}=\frac{m}{2R_1},\\
&\frac{4 p_0}{a^4}\frac{s^2+3}{s^4}=\frac{m}{2R_2},\\
&\frac{4 p_0}{a^2}\frac{s^2+1}{s^2}=m\delta_0. 
\end{aligned} \right. \label{eq:systspd}
\end{equation}
Dividing the first by the second, it turns out that the aspect ratio depends only on the initial geometry, since it solves
\begin{equation}
3s^4+\frac{R_1-R_2}{R_1}s^2-3\frac{R_2}{R_1}=0
\label{eq:fors4}
\end{equation}
via the only real positive root
\begin{equation}
s=\sqrt{\frac{R_2-R_1}{6R_1}+\sqrt{\frac{R_2}{R_1}+\left(\frac{R_2-R_1}{6R_1}\right)^2}}.
\label{eq:srat}
\end{equation}
Such solution is valid for any $R_1$ and $R_2$, if chosen according to Section \ref{sec:model}, including the eventuality that one of the two is negative, which is the common case of a contact between a concave and a convex bone extremity. Combining for instance the first and the third of the system \eqref{eq:systspd}, $\delta_0(t)$ and $p_0(t)$ are found as power functions of the semi-axis $a(t)$ as follows:
\begin{equation}
\delta_0(t)=\frac{s^2+1}{2R_1(3s^2+1)}a^2(t);\label{eq:de0}
\end{equation}
\begin{equation}
p_0(t)=\frac{m}{8 R_1} \frac{s^2}{3s^2+1}a^4(t).
\end{equation}
In view of the latter, Eq.~\eqref{eq:ppoft} becomes
\begin{equation}
p=\mathcal{G}P(x_1,x_2,t)=\frac{m}{8 R_1} \frac{s^2}{3s^2+1}\Psi(x_1,x_2,a(t))^2,
\label{eq:ppoft2}
\end{equation}
establishing that
\begin{equation}
\Psi(x_1,x_2,a(t))=a^2(t)-x_1^2-\frac{x_2^2}{s^2}.
\end{equation}

It remains to enforce the condition Eq.~\eqref{eq:equil} in order to gain the unknown $a(t)$. It is easy to integrate $p$ over $\omega(t)$ switching to elliptical coordinates with the result:
\begin{equation}
\iinteg{\omega(t)}{p(x_1,x_2,t)}{\omega}=\frac{m\pi s^3 }{24R_1(3s^2+1)}a^6(t).
\end{equation}

Recalling the definition of $p$ in Eq.~\eqref{eq:ppoft} and moving the time integral operator $\mathcal{G}$ out of the area integral, then the balance condition Eq.~\eqref{eq:equil} appears, leading to
\begin{equation}
a(t)=\left(\frac{24R_1(3s^2+1)}{m\pi s^3 }\mathcal{G}F(t)\right)^{1/6}.
\label{eq:at}
\end{equation}
In particular, the major semi-axis at the beginning of the loading $a_0=a(0)$ depends on the geometries and mechanical parameter $m$ of the two contacting bodies and the initial force $F_0=F(0)$ as
\begin{equation}
a_0=\left(\frac{24R_1(3s^2+1)}{m\pi s^3 }F_0\right)^{1/6}
\label{eq:a60}
\end{equation}
and allows to express $a(t)$, $A(t)$ and $\delta_0(t)$ more concisely as
\begin{equation}
\left(\frac{a(t)}{a_0}\right)^6=\left(\frac{A(t)}{A(0)}\right)^3=\left(\frac{\delta_0(t)}{\delta_0(0)}\right)^3=\mathcal{G}\frac{F(t)}{F_0}.\label{eq:Gconcise}
\end{equation}
The asymptotic solution Eq.\eqref{eq:disp_i} was obtained under the assumption that the loaded area size is much bigger than the layer thickness, thus $F_0$ can not be set to zero. The right-hand side of the latter equation results then never indeterminate.

In the case of time-independent coefficients $\alpha_i$ and $\beta_i$, the operator $\mathcal{G}$ can be inverted as next. Introducing the superscript $\sim$ to indicate the time Laplace transform of parameter $\sigma$, Eq.\eqref{eq:defG} yields to:
\begin{equation}\begin{split}
\frac{\tilde{y}}{\tilde{Y}}&=\left( 1+\sum_{i=1}^4 \frac{\alpha_i}{\sigma-\beta_i} +\frac{\alpha_5}{\sigma}\right)^{-1}=\sigma\frac{\mathcal{P}_n(\sigma^4)}{\mathcal{P}_d(\sigma^5)}=\sigma\sum_{i=1}^5\frac{B_i}{\sigma-\bar{\sigma_i}},
\end{split}\label{eq:laplapl}
\end{equation}
being $\bar{\sigma_i}$ and $B_i$ the poles and the residua of the polynomial fraction $\mathcal{P}_n/\mathcal{P}_d$. The remainder is surely zero because $\text{deg}\mathcal{P}_n<\text{deg}\mathcal{P}_d$. By applying the convolution theorem, the Laplace inversion of the latter gives
\begin{equation}
\mathcal{G}^{-1}Y(t)=y(t)=\sum_{i=1}^5B_iY(t)+\sum_{i=1}^5\bar{\sigma_i}B_i\dinteg{0}{t}{\me^{\bar{\sigma_i}(t-\theta)}Y(\theta)}{\theta}.\label{eq:Gminus1}
\end{equation}
With the inverse operator in the hand and after the due substitutions in Eq.~\eqref{eq:ppoft2}, finally the contact pressure can be obtained. Using the symbol $H$ for the Heaviside step function, 
for fulfilling the boundary conditions Eq.~\eqref{eq:BCs} also outside $\omega(t)$, one can write
\begin{equation}
P(x_1,x_2,t)=\frac{m s^2}{8R_1(3s^2+1)} \mathcal{G}^{-1}\Psi^2 H(\Psi),
\label{eq:Pall}
\end{equation}
where $H(\Psi)$ assumes the value 1 when $\Gamma(t)$ reaches the point of coordinates $(x_1,x_2)$. In the same way it is possible to trace back the individual surface displacements $w^{(i)}$ substituting
\begin{equation}
\Delta P(x_1,x_2,t)=m\mathcal{G}^{-1}(\Phi(x_1,x_2)-\delta_0(t))H(\Psi)
\end{equation}
coming from Eq.~\eqref{eq:cont2} into Eq.\eqref{eq:disp_i}. The problem stated in Section \ref{sec:model} results then analytically solved for the evolution of the contact domain and the bones approach as well as for the contact pressure distribution through Eqs.\eqref{eq:srat}, \eqref{eq:at}, \eqref{eq:de0} and \eqref{eq:Pall}.

\section{Numerical benchmarks}
\label{sec:numben}
Let us consider a single cartilagineous tissue to which the constitutive laws Eqs.~\eqref{eq:amatrix}~-~\eqref{eq:paramdef} apply. The TITH stiffness matrix components $A_{13}$, $A_{33}$, the only ones which contribute to the asymptotic solution in \cite{vitucci2016} together with $A_{44}$, can be rewritten as functions of the in-plane and out-of-plane Young's moduli $E_1$, $E_3$ and the Poisson ratios $\nu_1$, $\nu_{13}$ as follows:
\begin{equation}A_{13}=\frac{\nu_{13}}{1-\nu_1-2\nu_{13}^2\dfrac{E_3}{E_1}}E_3;\qquad A_{33}=\frac{1-\nu_1}{1-\nu_1-2\nu_{13}^2\dfrac{E_3}{E_1}}E_3.
\label{eq:As}\end{equation} 
The choice of the material parameters is not completely free though, but bounded by physical restrictions. Particularly, in order to preserve the solid matrix strain energy positivity, it was proved by \cite{auld1973acoustic} that, for TITH thin layers, it is required that
\begin{equation} 
A_{33}\geq A_{13},\qquad A_{33}\geq \frac{3}{4}A_{44}\geq 0.
\label{eq:auld}\end{equation}
Specific and separate values of $\nu_{13}$ and $\nu_1$ for a TITH cartilage layer have not been traditionally investigated, but the experimental studies which characterize the material as biphasic suggest that the apparent isotropic ratio is relatively small (e.g. see \cite{wang2003optical,keenan2009new,chegini2010time}). Therefore we assume for simplicity that $\nu_1=\nu_{13}=0$ within the next benchmarks. It is easy to show that, in such situation, Eqs.\eqref{eq:As}, combined with Auld's conditions Eq.\eqref{eq:auld}, shrink to
\begin{equation}
A_{13}=0,\qquad
A_{33}=E_3>0,\qquad
A_{33}\geq \frac{3}{4}A_{44} >0.
\label{eq:auld2}\end{equation}
\begin{figure}[htbp!]
\centering
\includegraphics[height=.45\textwidth]{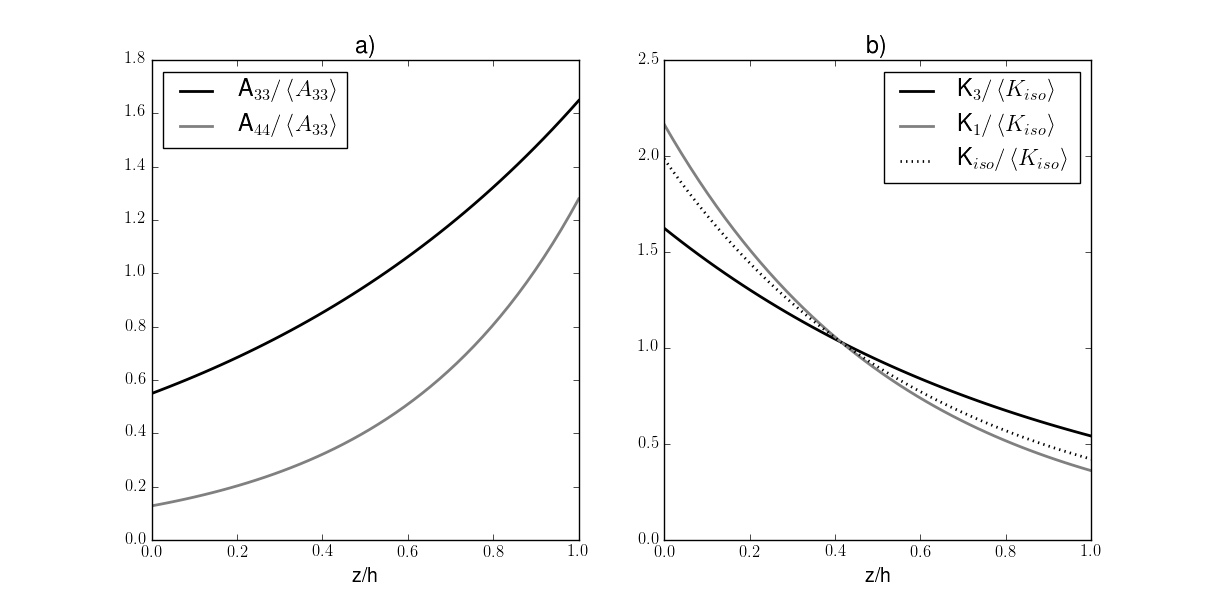}
\caption{In-depth distribution of material parameters. a) Stiffness matrix elements as multiples of $\Avg{A_{33}}$. b) Permeability components as multiples of $\Avg{K_{\text{iso}}}$.}
\label{fig:param2}
\end{figure}
At the same time, in \cite{wu2002elastic,federico2005transversely} it was shown the reason why a typical collagen distribution through the cartilage layer causes $E_3$ also to grow towards the tidemark at $z=1$, where $E_1>E_3$; vice versa $E_1$ decreases until it becomes smaller than $E_3$ at the tidemark. Since the proteoglycan gets more and more densely packed in along the thickness, the isotropic Young's modulus at the articular surface is smaller than at the bone attachment. The setting that we will use, which also accounts for these considerations, reads
\begin{equation}
A_{13}=0, \qquad \gamma=\frac{\log 3}{2},\qquad \alpha=\log 10.
\label{eq:apar}
\end{equation} 
The reader can notice, looking at Eq.\eqref{eq:paramdef}, that the shear modulus $A_{44}$ presents a tenfold increase through the depth similarly as in \cite{buckley2010high}, while the axial permeability $K_3$, linked to the axial stiffness inhomogeneity by the parameter $\gamma$, is let decrease three times toward the tidemark. As shown in \cite{federico2008anisotropy}, the planar permeability $K_1$ is expected to be larger than axial $K_3$ at the articular surface and vice versa at $z=0$ as the fluid flows easier along the prevailing collagen fibers orientation, while the overall equivalent isotropic $K_{\text{iso}}=\left(2K_1+K_3\right)/3$ steadily grows as a result of the decreased porosity. This leads us to the choice:
\begin{equation}
k_1=\frac{4}{3}k_{33},\qquad \gamma_1=\log 6.
\label{eq:kpar}
\end{equation}
The material was assigned average typical stiffness values $\Avg{A_{33}}=2\Avg{A_{44}}=0.5$MPa (e.g. see \cite{boschetti2004biomechanical}). Furthermore, an isotropic permeability was considered of average value $\Avg{K_{\text{iso}}}=2 \cdot 10^{-14}\text{m}^4\text{N}^{-1}\text{s}^{-1}$ similarly to the findings of \cite{boschetti2004biomechanical,boschetti2008mechanical}. The same properties are assigned to all the layers within the following benchmarks. The resulting distribution of the material parameters through the depth of the cartilage layer is shown in Fig.~\ref{fig:param2}.
\begin{figure}[ht]
\centering
\includegraphics[width=0.9\textwidth]{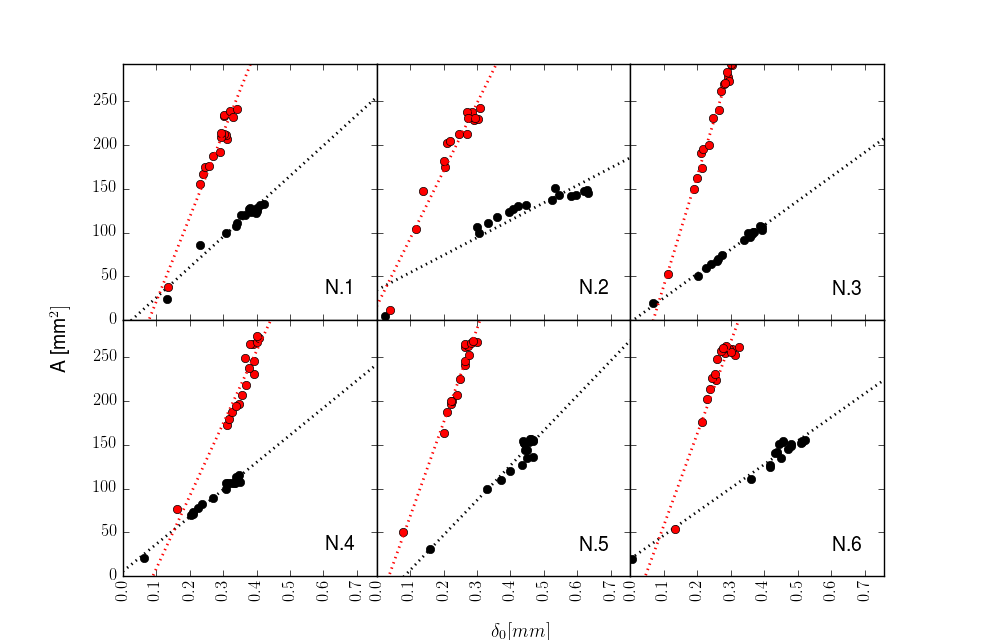}
\caption{Linear dependency of $A(t)$ and $\delta_0 (t)$ observed in the results published in \cite{hosseini2010vivo}. Six human tibiofemoral joints were loaded \textit{in vivo} and the contact area and bones approach were measured via MRI. Such dependency can be expained by Eq.\eqref{eq:adconst}. Points represent the experimental results, dotted lines their linear regression. Black indicates the lateral compartment, red the medial one.}\label{fig:linfit}
\end{figure}

\begin{figure}[ht]
\centering
\includegraphics[width=0.7\textwidth]{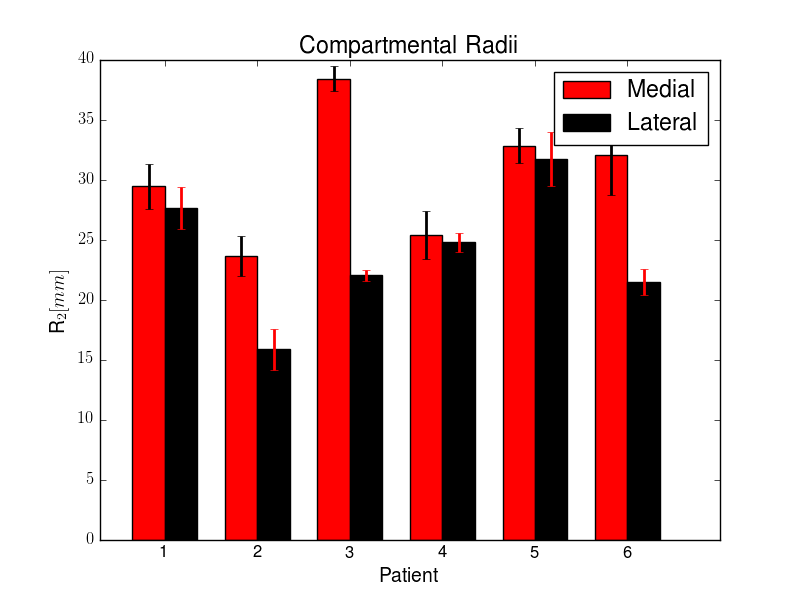}
\caption{Minimum contact curvature radii $R_2$ and respective error extracted from the experimental data published by \cite{hosseini2010vivo} assuming that the aspect ratios of the contact ellipses for the medial and lateral compartments of the tibiofemoral joint are $s^{(M)}=1$, $s^{(L)}=0.5$.}\label{fig:radii}
\end{figure}
Focusing on the tibiofemoral knee joint in extension, it is the locus where the two medial and lateral femoral condyles - respectively denoted $M$ and $L$ later on - contact the underlying tibial plateau. The latter is considerably flat at least in the stance contact area, which leads to choose the curvatures $1/{R_1^{(2)}}=1/{R_2^{(2)}}=0$. The medial condyle has been observed to be approximately spherical ((\cite{martelli2002shapes,kim2007world})) causing an approximately circular contact area, i.e. $R_{1}^{(1M)}/R_{2}^{(1M)}=1$, $s^{(M)}=1$. A visual estimate of the typical lateral contact area detected via magnetic resonance imaging (MRI) published in \cite{hosseini2010vivo} reveals a much tapered shape than in the medial compartment with an aspect ratio of about $s^{(L)}=0.5$ which indicates $R_{1}^{(1M)}/R_{2}^{(1M)}=7.43$ from Eq.~\eqref{eq:fors4}, where the reference axis $x_1$ is in the sagittal plane, $x_2$ in the coronal one. Making use of Eq.~\eqref{eq:de0}, one can deduce that the ratio
\begin{equation}
\frac{A(t)}{\delta_0(t)}=\frac{2\pi(3s^2+1)}{s^2+1}R_1
\label{eq:adconst}
\end{equation}
is supposed to be time independent according to our model. Analyzing the results of \cite{hosseini2010vivo} in terms of contact area and bones approach, the ratio ${A(t)}/{\delta_0(t)}$ presents indeed appreciably constant slope (see Fig.\ref{fig:linfit}), from which we are able to extract the unpublished size of their 6 patients for both the joint compartments expressed as contact radii. The retrieved values and standard deviations are illustrated in Fig.~\ref{fig:radii}. The average medial $R_2^{(1M)}$ and lateral $R_2^{(1L)}$ were found to be $30.3\pm 4.9$mm and $23.9\pm 5.0$mm with the same level of uncertainty. The minimum medial radius, in the coronal plane, resulted then approximately 27\% bigger than the lateral one, which is in good agreement with the findings of \cite{siebold2010computerized}. The fact that the maximum medial radius, the one in the sagittal plane, is much bigger and equal to $225.3\pm 36.5$mm agrees with the observation by \cite{martelli2002shapes}, where they noticed that, at stance, the medial condyle in such direction appears very flattened and a precise estimation of the contact radius results difficult. The two couple of average radii are adopted in the subsequent calculations together with the average thicknesses $h^{(M)}=1.3$mm and $h^{(L)}=1.6$mm extracted from \cite{hosseini2010vivo}.

\begin{figure}[ht]
\centering
\includegraphics[height=.45\textwidth]{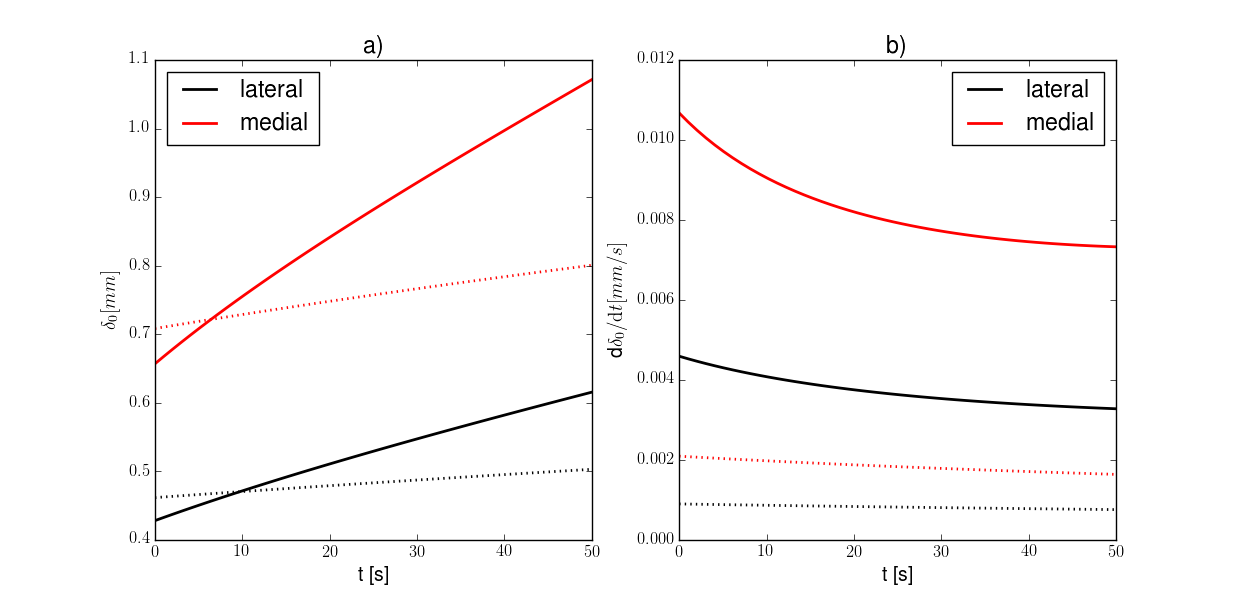}
\caption{First 50s of a constant load $F=700$N. Continuous lines illustrate the results for a TITH material, dotted ones indicate the isotropic homogeneous cartilage behaviour if averaged TITH stiffness and permeability are assigned. The medial compartment bears double as much load as the lateral one. a) Bones approach $\delta_0(t)$. b) Time-derivative of $\delta_0$.}\label{fig:coload}
\end{figure}
\begin{figure}[ht]
\centering
\includegraphics[height=.45\textwidth]{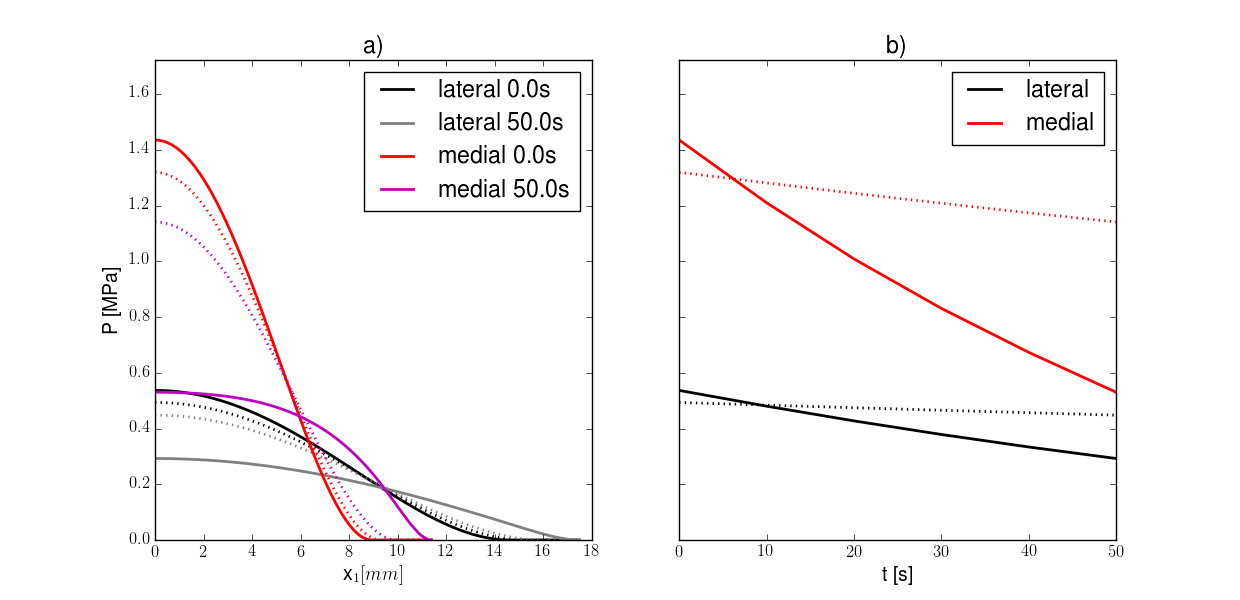}
\caption{Contact pressure distribution and evolution under constant load. Dotted lines stand for the isotropic homogeneous model. a) Distribution along the axis $x_1$ containing the major semiaxis $a(t)$ of the elliptical contact area soon after the loading and after 50s. b) First 50s of the pressure in the center of the contact area of coordinates $x_1=x_2=0$.}\label{fig:pconst}
\end{figure}
First we examine the case of a load deriving from a body weight of 700N at stance (about the European average according to \cite{walpole2012weight}), equally split between the two knees and distributed for $2/3$ and $1/3$ respectively on the medial and lateral compartments. It has been indeed measured that the medial compartment carries a much larger part of the load (\cite{werner2005effect,halder2012influence}). What we want to investigate is how an inhomogeneous distribution of stiffness and permeability may be able to improve the cartilage performance with respect to a homogeneous one whose properties present the same average across the thickness. In a way, how an actual cartilage arranges its mechanical resources for carrying out its functions. The resulting approach $\delta_0(t)$ in Fig.~\ref{fig:coload} is compared with the solution of \cite{argatov2011elliptical} for an isotropic homogeneous cartilage layer. For both the compartments, despite the initial value is smaller than according to such solution, $\delta_0$ grows remarkably quicker. Besides, its derivative, at least in first seconds, decreases much faster, pronouncing such desirable property already addressed in \cite{wu1997improved}. In Fig.~\ref{fig:pconst} we plot the contact pressure along the axis $x_1$ and its evolution in time. The results, in this case, exhibit not only a quantitative difference, but also a qualitative one. The curves in Fig.~\ref{fig:pconst}.a) for the TITH material do not deform homothetically during the expansion of the contact area like the homogeneous material would do, but the novel formula Eq.~\eqref{eq:Pall} allow them to change shape by flattening them at the origin. Here the mechanical convenience for the body in developing an inhomogeneous layer appears evident in the sense that, stated that one of the main functions of cartilage is to lower the pressure peaks, such aim is accomplished via a more even distribution of the force inside the contact area together with a faster decrease of the maximum pressure. It is in fact intuitive that it descends from the presence of a compliant zone of high permeability and low stiffness close to the cartilage surface.

\begin{figure}[!ht]
\centering
\includegraphics[height=.45\textwidth]{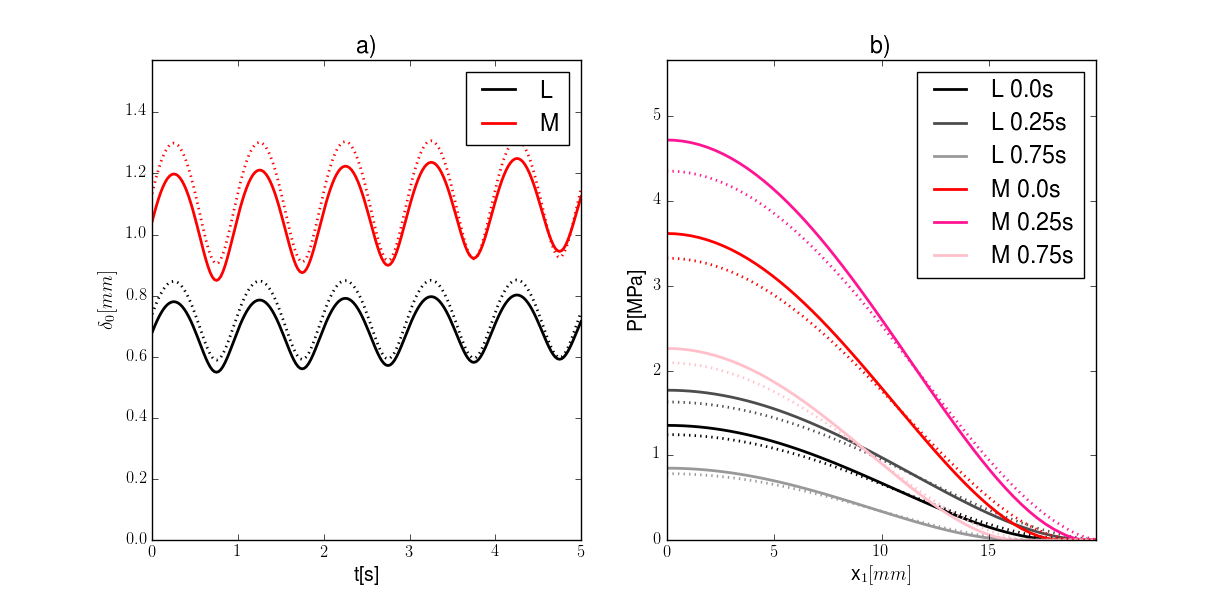}
\caption{Bones approach and consequent contact pressure profiles under oscillatinf load. Dotted lines show the the response of the isotropic homogeneous material averaged in the sense we discuss in the current section. a) First five cycles of deformations in therms of $\delta_0$. b) Contact pressure displayed at $t=0$ and the first maximum and minimum peak of deformation.}\label{fig:sinuso}
\end{figure}
The second load condition that we exemplify deals with a sinusoidal force of period 1s, similar to the frequency a normal human gait, and that oscillates on each knee between 0.5 and 1.5 of the same body weight of 700N. The portions absorbed by the two knee compartments stay the same as in the previous benchmark. Fig.~\ref{fig:sinuso}.a) shows the approach in the first five cycles for both compartments and compares it with the isotropic homogeneous solution homogenized as above. The oscillating part of $\delta_0(t)$ does not indicate a different behavior than the results obtained in \cite{argatov2011elliptical}, whereas it is clear that the steadier increase of the average trend is similar to the curves of Fig.~\ref{fig:coload}.a). In the short five cycles interval examined here, the effect discussed with regard to Fig.~\ref{fig:coload}.b) is not observable and in this case the difference between the continuous and dotted lines seems purely an outcome of the larger areas obtained for the isotropic homogeneous material with the particular homogenization criterion chosen in this Section. The oscillating deformation added onto a weakly increasing trend shows good agreement with the results of a similar load condition applied to two identical spherical homogeneous isotropic layers in \cite{wu2000joint}.

\section{Discussion}\label{sec:disc}
For the three-dimensional geometry described in Eq.\eqref{eq:parabi}, we were able to write the bones approach, the evolution of the contact area and the corresponding pressure distribution due to an arbitrary force applied onto the TITH biphasic cartilage layer treated in \cite{vitucci2016} (see Eqs.\eqref{eq:de0}, \eqref{eq:at}, \eqref{eq:Pall}). The solution is retrieved in closed-form and its exact within the assumptions of the model. The introduction of inhomogeneity and anisotropy allows to obtain a significantly different lowering of the peak contact pressure and growth of the contact area with respect to an isotropic homogeneous material whose properties are simply the average of the TITH one (see examples in Figs.~\ref{fig:coload},\ref{fig:pconst},\ref{fig:sinuso}). This proves once more that the scientist who intends to model the behavior of cartilage needs to pay a great attention to the interpretation of the material properties provided by experiments. The results are qualitatively similar to the analytical ones obtained by \cite{chidlow2013two} in the framework of functionally graded materials when dealing with an inhomogeneous elastic coating on top of an infinite half-space.

It seems remarkable that the ratio of the contact area and the bones approach is predicted to stay constant in time independently of the applied load as expressed in Eq.\eqref{eq:adconst}. By making use of the latter, together with assumptions on the ellipticity of the contact areas, it was possible to make very reasonable guesses about the originating contact radii which were not published in the work \cite{hosseini2010vivo}. The intercepts of the linear regressions in Fig.\ref{fig:linfit} were not zero though and it may derive from initial contact conditions that are different from the ones assumed by our model.

\begin{figure}[!ht]
\centering
\includegraphics[height=.5\textwidth]{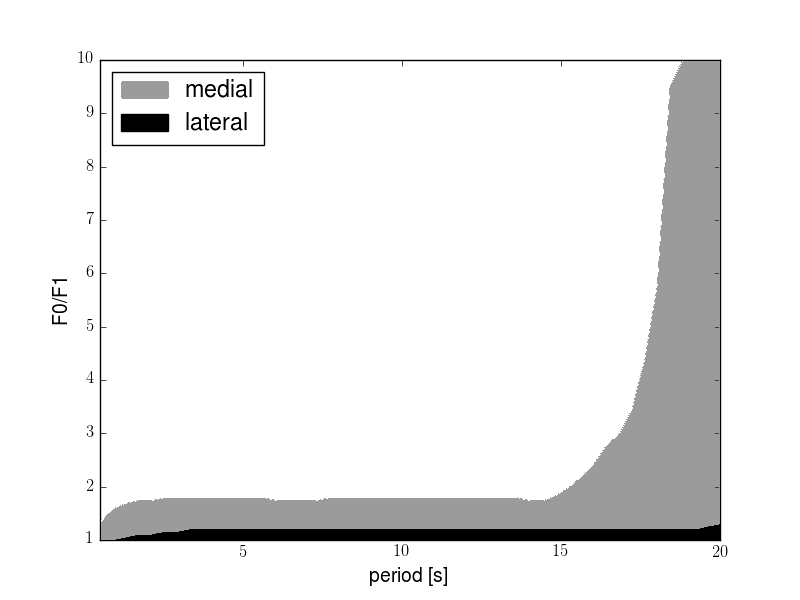}
\caption{Negative pressure arising under five cycles of loading for different ratios between the constant and oscillatory part $F_0$ and $F_1$ of the force and different oscillation periods.}\label{fig:negp}
\end{figure}
All the family of asymptotic solutions \cite{ateshian1994asymptotic,wu1996modeling,wu1997improved,argatov2011elliptical,argatovcontact,vitucci2016}
do not take into account the dependency of the permeability on the volumetric strain that has been well known since the study \cite{mow1980biphasic}. If this simplification of the equations permits the advantageous feature of closed-form, easily analyzable solutions, on the other hand it causes the deformation to emerge unbounded. It can be seen, for instance, in Fig.\ref{fig:negp} that negative pressure may arise due to such unboundedness for high oscillating load portions and always for prolonged load application, visible on the right-hand side of the plot. Such values of the pressure are obviously unphysical and not acceptable, given that no adhesion is assigned to the layers surfaces. On the other hand, that the validity of the proposed approach is constrained to short time response is part of the model preconditions. Asymptotic formulas of the kind of Eq.\eqref{eq:disp_i} are in fact reliable under the assumption that the considered time is much smaller than the hydrogel characteristic time $\tau_{\text{gel}}=h^2/(A_{33}K_3)$ which takes the value of about 290s in the benchmarks of Sec.\ref{sec:numben}. An asymptotic solution which includes the effect of strain-dependent permeability is currently under investigation and would presumably make the present model applicable also for later times. 

The advances in imaging techniques allow nowadays to obtain very detailed measurements from \textit{in vivo} experiments on articular cartilage. A considerable amount of studies have been published on the topic, among which \cite{herberhold1999situ,song2006articular,wan2008vivo,li2008determination,bingham2008vivo,hosseini2010vivo,shin2011vivo,chan2016vivo}. The field looks ready for enhancing early diagnoses of degenerative pathologies such as osteoarthritis with consequent benefits for the patients. In order to exploit such technological advantages, though, more extended studies need to be conducted. It would be seriously fruitful for the scientists working on mechanical modeling to see more data provided in these publications regarding at least the geometry of the contact areas and of the contact surfaces, the forces applied onto the single articular cartilages. This way a correct modeling could finally lead to real time analyses and standard procedures.


\section*{Acknowledgments}
Gennaro Vitucci and Gennady Mishuris participated in this work under the support of the European projects, respectively, \textit{FP7-MC-ITN-2013-606878-CERMAT2} and \textit{PIRSES-GA-2013-610547-TAMER}. The authors feel to warmly acknowledge Ivan Argatov for all the interesting and fruitful discussions. 

\bibliographystyle{abbrv}
\bibliography{Bibliography}



\end{document}